\documentstyle[prl,aps]{revtex}
\begin{document}
\bibliographystyle{unsrt}
\title{Superfluid Spin-down, with Random Unpinning of
the Vortices}
\author{M. Jahan-Miri$^*$}
\address{Department of Physics, Shiraz University, Shiraz 71454, Iran}

\date{\today}
\maketitle
\begin{abstract}
The so-called ``creeping'' motion of the pinned vortices in a
rotating superfluid involves ``random unpinning'' and ``vortex
motion'' as two physically separate processes. We argue that such
a creeping motion of the vortices need not be (biased) in the
direction of an existing radial Magnus force, nor should a
constant microscopic radial velocity be assigned to the vortex
motion, in contradiction with the basic assumptions of the
``vortex creep'' model. We point out internal inconsistencies in
the predictions of this model which arise due to this unjustified
foundation that ignores the role of the actual torque on the
superfluid. The proper spin-down rate of a pinned superfluid is
then calculated and turns out to be much less than that suggested
in the vortex creep model, hence being of even less observational
significance for its possible application in explaining the
post-glitch relaxations of the radio pulsars.
\end{abstract}


\section{Introduction}
Spinning down (up) of a superfluid at a given rate is associated
with a corresponding rate of outward (inward) radial motion of its
vortices. If the vortices are subject to pinning, as it is assumed
for the superfluid in the crust of a neutron star, a spin-down
would require unpinning of the vortices (from the lattice nuclei).
The model of ``vortex creep" \cite{alet84} envisages such a
spinning down to occur through quantum tunnelling {\it between}
adjacent pinning {\it sites}. Here, we aim to show that the model
is internally inconsistent, and also contradicts the well-known
general requirements for a superfluid spin-down process. We argue
that, while tunnelling and/or thermal activation could only help
the vortices to overcome the pinning barriers, however any
possible (radial) motion of the vortices (before repinning) is a
separate {\em dynamical} process, subject to their equation of
motion. The vortex radial motion is, as in the absence of any
pinning, determined by the {\it external} torque on the superfluid
(as a whole including its vortices). The same torque might as well
be termed {\em internal} with respect to the superfluid and its
container, but should be distinguished from the external torque on
the {\em container}, as well as the internal torque between the
superfluid and its vortices. As is well-known, the external torque
on a superfluid is primarily exerted on the vortex cores, and may
be realized {\it only} when the vortices (tend to) have an
instantaneous azimuthal velocity relative to the ``container''
(ie. the crust of the star in this case, consisting of the solid
lattice, phonons, and the permeating electron gas). Thus, in the
presence of pinning, the vortices may or may not tend to undergo a
radial displacement upon unpinning (as for the free vortices in
the absence of any pinning); an unpinned vortex might as well {\em
repin at the same site}. In contrast, the vortex creep model
implies the following equivalence:
\begin{eqnarray}
vortex \  unpinning \stackrel{?}{\equiv} vortex \  radial \ motion.
\end{eqnarray}
in the sense that both are related to a single cause, in that
model. It should be however apparent that the two are physically
distinct processes which in principle may or may not occur
simultaneously. In fact, the misconception and mixing of the two
processes is inherent in the adopted terminology, ie. ``creeping
vortices'', which may be replaced by a ``random unpinning''
followed by a ``vortex motion'' before repinning. ``Creeping'' may
be indeed saved for the case of vortices (fluxoids) in a
superconductor where their (radial) displacements bear no
dynamical significance, as such. Clearly, a quantum tunnelling of
the vortices between pinning sites at different radial positions
(when realized) would involve both the processes simultaneously.
Nevertheless, the combined processes would amount to a transition
between states with different angular momenta, hence different
energies. To invoke such a quantum description of the phenomenon,
however, one needs to work out the problem of superfluid spin-down
(-up) self-consistently in quantum mechanics, at least
qualitatively. The relevant transition rules have to be taken into
account, while allowing for a transfer of angular momentum (which
is associated with a radial motion of the vortices) between the
superfluid and the container/normal fluid. The analogy with the
tunnelling of a particle out of a potential well is not
straightforward in the case of ``hopping'' of a vortex if a {\em
radial} displacement is also involved. A proper consideration for
the $r$-dependence of the angular momentum carried by a vortex
should be made, in order to describe a vortex radial motion in
terms of a quantum tunnelling (transition). In contrast, the {\em
unpinning} event by itself, without any implicitly assumed radial
displacement of the unpinned vortex, might provide a
straightforward analogy with the case of particle in a potential
well. A quantum mechanical treatment of the spin-down process has
not been, so far, addressed in the context of the vortex creep
model, and we would likewise adhere to the classical
hydrodynamical description of the vortex motion as is commonly
adopted \cite{son87}.

It should be also emphasized that the original prescription of the
vortex creep model for the spin-down rate of a pinned superfluid
\cite{alet84} has remained the same throughout all the subsequent
applications and modifications of the model \cite{accp96}. The
feature which has been changing is the assumed detailed and
complicated picture of the numerous superfluid layers within the
crust of a neutron star, having different pinning and unpinning
properties. We will be however concerned only with the basic
relation suggested in that model for the spin-down rate of a
pinned superfluid. More sophisticated treatments of the vortex
creep process have also appeared \cite{jonAp91,elb92} discussing
the physics of pinning/unpinning in great depths and details.
Nevertheless the dynamical significance of the vortex radial
motion does not seem to have been emphasized and treated properly,
in this context. Moreover, the spin-down (-up) process discussed
here is a general phenomenon of the pinned superfluidity,
applicable to the laboratory experiments on superfluid Helium
\cite{TT,hed80,sch81,ziv2}, as well, in addition to its common
application in the case of neutron stars. The predicted rate may
be indeed tested experimentally, at least in principle. Even
though the process has not been so far invoked in this context,
however there exist no fundamental reasons against its potential
application in future experiments.

In section 2 the well-known physics of superfluid spin-down (up)
is stated briefly, highlighting the prime role played by the
vortices in communicating the external torque to the bulk
superfluid. It is then pointed out that the vortex creep
formulation does not incorporate this fundamental role played by
the vortices. Section 3 contains our reasoning against  the model,
in further details. Two misleading assumptions in the model are
discussed in the two subsections separately. In section 4 a
revised form of the superfluid spin-down rate, in the presence of
pinning, is suggested, depending on the two possibilities
discussed for the relative azimuthal motion of the unpinned
vortices.

\section{Spin-down and Vortex Creep}
Superfluid vortices move with the local superfluid velocity except
when there is an external force acting on them (see Eqs 3 \& 6
below). A torque on the superfluid, acting primarily on the
vortices, results in a vortex radial velocity $v_r$ corresponding
to a given rate $\dot \Omega_{\rm s}$ of change of the rotation
frequency $\Omega_{\rm s}$ of the superfluid:
\begin{eqnarray}
v_r &=& -{ r \over 2}{{\dot \Omega_{\rm s}}\over \Omega_{\rm s}},
\end{eqnarray}
where $r$ is the distance from rotation axis, and $v_r>0$ is in
the outward direction. In the case of pinned vortices the
superfluid spin-down, or equivalently the vortex radial motion, is
subject to unpinning of the vortices (at least temporarily). The
required unpinning of the vortices might be achieved under the
influence of a Magnus force $\vec{F}_{\rm M}$ acting on the
vortices, which is given, per unit length, as \cite{son87}
\begin{eqnarray}
\vec{F}_{\rm M} & = & -  \rho_{\rm s} \vec{\kappa} \times
                        (\vec{v}_{\rm s} - \vec{v}_{\rm L}),
\end{eqnarray}
where $\rho_{\rm s}$ is the superfluid density, $\vec{\kappa}$ is
the vorticity of the vortex line directed along the rotation axis,
$\vec{v_{\rm s}}$ is the local velocity of the superfluid, and
$\vec{v_{\rm L}}$ is the velocity of a vortex-line. Accordingly,
if a lag \( \omega \equiv \Omega_{\rm s} - \Omega_{\rm c}\) exists
between the rotation frequency of the superfluid and that of the
vortices (pinned and co-rotating with the crust at $\Omega_{\rm
c}$ ) a radially directed Magnus force \((F_{\rm M})_r= \rho_{\rm
s} \kappa r \omega \) would act on the vortices, where \(\omega >0
\) corresponds to an outward directed $(F_{\rm M})_r$, vice-versa.
A pinned superfluid may therefore follow the steady-state spinning
down of its container provided that $\omega \geq \omega_{\rm
crit}$, where $\omega_{\rm crit}$ is the critical lag value
required for the Magnus force $(F_{\rm M})_r$ to overcome the
pinning forces. Likewise, any other mechanism for an unpinning of
the vortices (say, their random unpinning through quantum
tunnelling) would have a role similar (and added) to that of the
Magnus force above. The important point to note is that, whatever
the unpinning mechanism might be, the basic role played by the
vortices in transferring a torque to the superfluid could not be
different from that in the absence of any pinning, which is well
established \cite{son87,titi90}.

In contrast, the existing formulation of the vortex creep model is
such that if a superfluid spin-down were to be ``driven'' by the
process of unpinning itself. The evidence is that the suggested
spin-down rate is independent of the instantaneous value of the
external torque on the superfluid. The fundamental relation in
that model for the vortex radial velocity is (Eq.~17 in Ref.~[1]):
\begin{eqnarray}
v_r = v_0 \exp{[- \frac{E_{\rm p}}{{\rm k}T}
\frac{\omega_{\rm cr}-\omega}{\omega_{\rm cr}}]},
\end{eqnarray}
where $E_{\rm p}$ is the pinning energy, k is the Boltzmann
constant, $T$ is the temperature, and $v_0$ is {\it a constant}
referred to as the ``typical microscopic velocity'' of the
unpinned vortices. The relation (Eq.~4) consists of two terms. The
exponential term (arising from terms $e^{-\Delta E/kT}$, where
$\Delta E$ is the relevant energy barrier; see Eqs~14--16 in
Ref.~[1]) represents the rate coefficient for the unpinning
events. That is, the probability of a vortex being free, in close
analogy with the case of unpinning of fluxoids in a hard
superconductor \cite{ank64}. The other term, ie. $v_0$, is an
assumed {\em constant} velocity for the radial motion of the
vortices. Thus, the averaged radial velocity of the vortices as
given by Eq.~4, hence the corresponding superfluid spin-down rate
(Eq.~2), are obviously independent of the presence or absence of
an external torque on the superfluid. We caution again that the
fundamental quantity missing in this formula, ie. the actual
external torque on the superfluid, should not be confused with the
external torque on the container which is further introduced in
the formulation of the creep model.

However, Eq.~4 is further transformed, in the creep model, into
the following form which artificially introduces the missing role
of the actual torque. This is achieved by simultaneously solving
the above equation together with the one governing the rotational
dynamics of the whole system of the superfluid plus the container,
which results in (Eq.~28 in Ref.~[1])
\begin{eqnarray}
v_r = v_{\infty} \exp{[- \frac{E_{\rm p}}{{\rm k}T}
\frac{\omega_{\infty} - \omega}{\omega_{\rm cr}}]},
\end{eqnarray}
where $v_{\infty}$ is the average radial velocity corresponding to
the steady state spin-down rate $\dot\Omega_{\infty}=\frac{N_{\rm
ext}}{I}$ of the superfluid along with the crust in response to
the external torque $N_{\rm ext}$ acting on the crust (the
superfluid container), $I$ is the total moment of inertia of the
system, and $\omega_{\infty}$ is the steady-state value of the lag
defined as the value of $\omega$ in Eq.~4 for which $v_r=
v_{\infty}$. Note that the sign of $v_r$ in Eq.~4 is to be decided
by the sign of $\omega$, however in Eq.~5 it is not clear how the
sign should be determined since $v_{\infty}$ might impose a sign
for $v_r$ opposite to that required by $\omega$; see item two
below. The illusive role of the external torque (even) on the
container, represented  by $v_\infty$ in Eq.~5, and the internal
inconsistency of the model may be further appreciated from the
following contradictory points.
\begin{itemize}
\item If $N_{\rm ext} =0$, ie. in the absence of an external
torque acting on the container of a pinned superfluid, Eq.~4
implies $\dot\Omega_{\rm s} \neq 0$ while Eq.~5 results in
$\dot\Omega_{\rm s} = 0$. \item Also one might consider the
initial condition $\omega < 0$, in the presence of an external
torque $N_{\rm ext}<0$, a case which has been specifically
suggested, in the vortex creep, to explain the large post-glitch
spin-down rates observed in some pulsars \cite{apc90}. Again,
Eqs~4 \& 5 have contradictory predictions of $\dot\Omega_{\rm s} >
0$ (because of the inward radial Magnus force, hence the inward
bias of creeping motions) and $\dot\Omega_{\rm s} < 0$ (due to the
sign of $N_{\rm ext}$, hence the sign of $v_r$), respectively. The
artificial appearance of $N_{\rm ext}$ (through $v_\infty$ in
Eq.~5) in the model may be best observed from the fact that indeed
a spin-{\em up} ($\dot\Omega_{\rm s} > 0$) has been favored in the
vortex creep for this case \cite{apc90}.
\end{itemize}
We note that the above contradictions persist even for the more
general form of Eq.~4 (ie. Eq.~16 in Ref.~[1]), with a $sinh$
dependence instead of the exponential term in Eq.~4; the
exponential or the $sinh$ terms have no influence on the sign of
$v_r$. The above contradictory predictions of the vortex creep
model is rather due to the fact that their original relation for
the spin-down rate of a pinned superfluid (Eq.~4) makes by itself
a definite prediction; there exist no free parameter to be further
fixed when it is solved together with the equation governing the
dynamics of the whole system. Hence Eq.~5 which is supposed to
incorporate also the effect of the external torque is inconsistent
with Eq.~4.

\section{Main Objections}
The missing role of a torque, between the superfluid and its {\em
container}, is imitated in the model by virtue of the following
two unjustified assumptions, which artificially compensate for the
sign and the magnitude of the torque, respectively. Firstly, Eq.~4
is derived assuming that a {\it radial} Magnus force causes a {\it
radial} bias in the (creeping) motion of vortices. Moreover, a
{\it constant} radial velocity $v_0$ (Eq.~4) has been assigned to
the unpinned moving vortices, irrespective of the actual
instantaneous torque that may or may not be realized. A detailed
discussion of the two points will follow.

\subsection{Radial Bias}  The derivation of Eq.~4 in the vortex
creep model is based on the assumption that random creeping motion
of pinned vortices would be biased in the, say, outward radial
direction if there exists a Magnus force acting in the same radial
direction. This has been further argued to be a consequence of the
presence of a slope (a bias) in the radial profile of free energy
of the vortices (see Fig.~3 in Ref.~[1], and also Eq.~5 in
Ref.~[5]). It is however noted that the slope in the potential
energy is realized {\em only if} vortices are already moving
radially \cite{shah80}. The slope, arising from the $r$-dependence
of the Magnus force (Eq.~3), does not have the usual dynamical
interpretation implying a ``down-the-hill'' motion. In other
words, the slope is a consequence of, {\em not} a cause for, the
motion. This unusual dynamical behavior is of course another
aspect of the basic property of the vortices that move in a
direction perpendicular to an applied force. It might be
instructive to note that the original formulation of the flux
creep in hard superconductors had also been criticized on the same
footing even though vortices do not play any dynamical role in
that context \cite{vin64}. In short, the derivation of Eq.~4 is
unjustified for the obvious reason that a {\em radial} Magnus
force should not be associated with a {\em radial} vortex motion.

Furthermore, the (radial) Magnus force could not be the primary
cause of the radial motion of the vortices also because it is an
{\em internal} force exerted by the superfluid on the vortices and
thus could not be the source of a torque on the fluid itself. The
role of the Magnus force is rather to assist vortices to overcome
the pinning barriers, {\em independent of its direction}. That is,
it drives the system towards a state similar to the absence of any
pinning, by increasing the instantaneous number of free vortices.
The subsequent motion of the unpinned vortices until their
repinning is decided by the vortex equation of motion
\cite{son87}:
\begin{eqnarray}
        \vec{F}_{\rm ext} + \vec{F}_{\rm M}=0
\end{eqnarray}
where $\vec{F}_{\rm ext}$ is the external force on a vortex, per
unit length, exerted by the environment of the superfluid (ie. by
the crust). Any radial motion of the vortices, requires (Eqs 3 \&
6) a corresponding {\em azimuthal external} force $F_{\rm ext}$
acting on the moving vortices, instantaneously.

\subsection{Constant $v_0$}  The vortex creep model also
introduces a constant ``microscopic'' velocity $v_0$, for the
vortex radial motion (Eq.~4). Firstly, the term ``microscopic''
might be misleading, because here one may not think in terms of a
microscopic velocity as opposed to a (macroscopic) drift velocity
observed in, say, statistical mechanics of gas particles.
Superfluid vortices do not obey Newtonian dynamics (or its
equivalent formulations) upon which the common notions of the
statistical mechanics lie. Thus, unlike ordinary gas particles
which have constant microscopic thermal velocities at a given
temperature, even in the absence of external forces, a vortex may
move with a constant velocity, be it called microscopic or
macroscopic, only as long as an {\em external} force acts on it,
as required by Eq.~6. The above notion of ``microscopic velocity''
should be interpreted accordingly. Thence, equation 4 may be
readily disqualified, given that the other exponential (or $sinh$)
term therein accounts only for the vortex unpinning probability.
The obvious reason is that, it assigns a constant radial velocity
to the unpinned vortices, irrespective of the presence and the
magnitude of the external azimuthal forces which should act on the
vortex cores for a spin-down to be achieved. The required torque
for a change in the spin frequency of the superfluid is determined
by the existing forces, which would equivalently determine (Eqs~3
\& 6) the magnitude of the (microscopic) radial velocity of
instantaneously moving vortices. The radial motion of vortices
during a change in the spin frequency of a superfluid has to be
accompanied by a corresponding azimuthal one (in their
instantaneous rotating frame). This may be seen directly from a
solution of the equation of motion of vortices during a superfluid
rotational relaxation (Eq.~9 in Ref.~[17], and Eq.~4 in
Ref.~[18]). A torque may be transmitted to a superfluid only by
virtue of the simultaneous and corresponding azimuthal-radial
motions of the vortices. The radial motion is initiated by the
azimuthal force on vortices, which in turn relies on the vortex
azimuthal motion relative to the environment of the superfluid.
The constant $v_0$, as in Eq.~4, does not comply to these
well-known requirements of the vortex dynamics.

Curiously, no explicit derivation for the constant velocity, and
its magnitude $v_0=10^7 {\rm cm \ s}^{-1}$, may be found in the
related published literature, in spite of its prime significance
in determining the rate of spinning down of a superfluid. There
exists only a short comment indicating that it is associated with
the so-called ``Bernoulli forces'' \cite{alet84}. Such a force is
defined \cite{shah80} to be a generalization of the Magnus force
in presence of superfluid density variations (between the interior
and exterior of the nuclei). The same force, acting as a repulsive
central force field between a vortex and the nuclei (ie. pinning
centers), is in fact invoked as part of the pinning forces which
prevent a superfluid to respond to an otherwise active torque! The
definition \cite{shah80} is, nevertheless, for an assumed {\it
cylindrical} geometry of a nucleus, and its generalization to the
real spherical case of pinning sites (the nuclei) must be unknown,
as is the case with the configuration of vortex lattice inside a
spherical container \cite{titi90}. Hence, the way to a
quantitative estimate of the force, thus $v_0$, must be obscured.
Notwithstanding, whatever the magnitude of the assumed force might
be, the resulting radial velocity could not be a constant.

In the absence of any definite supporting argument, a discussion
of the irrelevance of the Bernoulli force to the torque on
superfluid would be naturally ambiguous and superficial. Some
general remarks might be made, however. A vortex will be subject
to the similar central force fields, around its successive
unpinning and repinning sites, which would act in azimuthal {\it
opposite} directions; the two effects might as well cancel out
with no net torque being imparted. Moreover, one may ask how the
same force could be simultaneously responsible for a {\it
stationary pinning} condition as well as the {\it torque}. Notice
that the vortex is assumed to escape the pinning barrier through a
quantum tunnelling, thus the pinning potential does not do any
work on the vortex, by definition. Finally, since the assumed
force arises from a Magnus effect one might doubt whether it is
imparted by the nuclei, or by the superfluid itself in which case
it could not be the source of a torque on the superfluid.

The external forces on vortices could, in general, be of a viscous
drag or a ``static'' frictional nature \cite{adam85,jon91}. The
latter type, associated with the ``pinning'' forces should not be
however confused with the role of pinning forces on the pinned
vortices co-rotating with the pinning centers. In order for the
pinning forces to act as frictional forces and impart a net torque
on the superfluid the vortices should remain unpinned due to the
effect of Magnus force \cite{adam85}. This requires
$|\omega|>\omega_{\rm crit}$ which means there should be no
stationary pinning, hence no ``creeping,'' to start with.
Therefore, the static frictional forces are not relevant to the
cases of interest here, given $|\omega|<\omega_{\rm crit}$.

\section{Estimating the Spin-down Rate}
A proper formulation of the superfluid spin-down rate in presence
of the random unpinning is however straightforward. The
corresponding averaged radial velocity $v_r$ of the vortices is
determined by the unpinning probability for each vortex times the
radial velocity $v_{\rm f}$ of an unpinned vortex during its
motion until repinning. That is,
\begin{eqnarray}
v_r = P_{\rm u} \  v_{\rm f}
\end{eqnarray}
where $P_{\rm u}$ is the unpinning probability , that is the
weight function for the instantaneous number of (unpinned) moving
vortices. The unpinning probability, corresponding to the energy
barrier $\Delta E= E_{\rm p}(1- \omega/\omega_{\rm cr})$, is given
as \cite{ank64,alet84}
\begin{eqnarray}
P_{\rm u}= \exp{(-\Delta E/ kT)}= \exp{[- \frac{E_{\rm p}}{{\rm
k}T} \frac{\omega_{\rm cr}-\omega}{\omega_{\rm cr}}]}
\end{eqnarray}
Substituting in Eq.~7, the revised spin-down rate of a pinned
superfluid is derived as
\begin{eqnarray}
v_r = v_{\rm f} \exp{[- \frac{E_{\rm p}}{{\rm k}T}
\frac{\omega_{\rm cr}-\omega}{\omega_{\rm cr}}]}.
\end{eqnarray}
In contrast to the constant $v_0$ in the earlier rate given by
Eq.~4, both the magnitude and the (inward or outward radial)
direction of $v_{\rm f}$, in Eq.~9, are to be determined by the
instantaneous azimuthal external force $F_{\rm ext}$ (Eq.~6)
acting on the moving vortices. The external force, being the
viscous drag of the permeating electron (and phonon) gas
co-rotating with the crust, depends on the relative azimuthal
velocity $v_{\rm rel}$ between the {\it crust} and the {\it
unpinned vortices}, and on the associated velocity-relaxation
timescale $\tau_v$ of the vortices. The drag force, per unit
length, is given as \cite{as88,mj98}
\begin{eqnarray}
n_{\rm v} F_{\rm ext} & = &  \rho_{\rm c} {v_{\rm rel}
\over \tau_v },
\end{eqnarray}
where $n_{\rm v}$ is the number density of the vortices per unit
area, and $\rho_{\rm c}$ is the effective density of the
``crust''. For a given force on the vortices, Eq.~10 might be
interpreted as the defining relation for $\tau_v$, which is the
velocity-relaxation timescale of the microscopic constituents of
the system (as for, say, particles in a normal gas).

In order to determine $v_{\rm rel}$ one might distinguish between
two distinct possibilities, which has not been addressed
previously. When a vortex becomes unpinned it might be expected to
either
\begin{itemize}
\item[{\em i)}] maintain its overall co-rotation with the pinned
vortex lattice; hence
\begin{eqnarray}
v_{\rm rel} \sim r \ \frac{I}{I-I_{\rm s}} \ \tau_{\rm D} \
\dot\Omega_\infty,
\end{eqnarray}
due to the spinning down of the crust, at a rate
$\dot\Omega_\infty$, under the influence of an external torque,
where $\tau_{\rm D}$ is the superfluid dynamical coupling
timescale (see below), $I_{\rm s}$ is the moment of inertia of the
superfluid, and $I$ is that of the superfluid plus its container
(ie. the whole star). Else, the unpinned vortex may \item[ {\em
ii)}] jump instantaneously to a rotation frequency \(\Omega_{\rm
L}=\Omega_{\rm s}\); hence
\begin{eqnarray}
v_{\rm rel} \sim r \ \omega.
\end{eqnarray}
\end{itemize}
Case {\em (i)} could arise due to the general requirement for a
locally uniform vortex distribution imposed by the minimization of
the free energy. Case {\em (ii)}, on the other hand, might be
realized because of (Eq.~6) the presence of an otherwise
unbalanced {\em radial} force $(F_{\rm M})_r$ on an unpinned
vortex, also considering the usual approximation of zero inertial
mass for a vortex \cite{son87,bach83}. Either of the two
possibilities might provide a better approximation depending on
whether a vortex unpin as a whole along its length (Case {\em
ii}), or only small segments of it are unpinned randomly. Case
{\em (i)} is probably more favorable for the superfluid in the
crust of neutron stars, given the huge number of the pinning
centers along each vortex which prevent a complete unpinning of
the whole vortex. In contrast, for laboratory experiments where
only the end point(s) of a vortex is pinned the latter case might
be relevant.

In spite of the above uncertainties, a tentative estimate of the
magnitude of $v_{\rm f}$, for each of the two cases, might be
instructive. For this purpose, we consider the limiting situation
with an unpinning probability $P_{\rm u} \lesssim 1$. Thus $v_{\rm
f}=v_r$, and from Eq.~2
\begin{eqnarray}
v_{\rm f} &=& -{ r \over 2}{{\dot \Omega_{\rm s}}\over \Omega_{\rm
s}},
\end{eqnarray}
On the other hand, for an assumed two-component model of the
superfluid plus the ``crust'', the superfluid spin-down rate may
be given as \cite{Bayet69}
\begin{eqnarray}
\dot\Omega_{\rm s} \sim \frac{I-I_{\rm s}}{I} \frac{v_{\rm rel}}{r
\tau_{\rm D}},
\end{eqnarray}
in accord with the estimate given in Eq.~11.

The relation between $\tau_{\rm D}$ and the vortex velocity
relaxation timescale, $\tau_v$, may be determined from a solution
of the vortex equation of motion (Eq.~6). This will indeed secure
the crucial dependence of the microscopic velocity $v_{\rm f}$ of
the unpinned vortices on the instantaneous external forces
($F_{\rm ext}$) acting on them; the vital dependence that is
missing in the creep model. The quantity $\tau_{\rm D}$ is the
macroscopic timescale for the dynamical coupling of the assumed
two-component model of the superfluid and its container (ie. the
rest of a neutron star apart from the superfluid). In other words,
$\tau_{\rm D}$ is the time needed for the simultaneous
readjustment of the vortices as a whole, in both the radial and
azimuthal directions, in response to a given torque on the
superfluid. The calculated general relation between $\tau_{\rm D}$
and $\tau_v$ \cite{adam85,mj98} gives the following approximate
relation:
\begin{eqnarray}
\tau_{\rm D} \sim {{I_{\rm s}} \over {I}}\tau_v.
\end{eqnarray}
Substituting for $ \tau_{\rm D}$ (Eq.~15) and $\dot\Omega_{\rm s}$
(Eq.~14) back in Eq.~13 one derives
\begin{eqnarray}
v_{\rm f} \sim \frac{(I-I_{\rm s})}{I_{\rm s}} \frac{v_{\rm
rel}}{2 \Omega_{\rm s} \tau_v}.
\end{eqnarray}
Finally, either of the two estimates given for $v_{\rm rel}$
(Eq.~11 or 12), may be substituted for to derive
\begin{eqnarray}
     v_{\rm f} \sim  \begin{array}{ll}
                \frac{r}{2 \Omega_{\rm s}} \dot\Omega_\infty &
                \ \ \ \ \ \mbox{case ($i$)} \\
                             \frac{r}{2 \Omega_{\rm s}}
         \frac{(I-I_{\rm s})}{I_{\rm s}} \frac{\omega}{\tau_v} &
               \ \ \ \ \  \mbox{case ($ii$),}
                            \end{array}
\end{eqnarray}

A comparison of the corresponding spin-down rate of the pinned
superfluid in the crust of a neutron star with that predicted in
the creep model would be instructive. Using typical values of the
parameters such as $r\sim10^6 \ {\rm cm}$, $\Omega_{\rm s} \sim
10^2 {\rm rad~s}^{-1}$, $I_{\rm s}/I \sim 0.02$,
$\dot\Omega_\infty \sim 10^{-10}  {\rm rad~s}^{-2}$, $\omega \sim
10^{-2} {\rm rad~s}^{-1}$, $\tau_v \sim 10^{-1} {\rm s}$, one
obtains \( v_{\rm f} \sim 10^{-5}, {\rm or} 10^{3} \ {\rm cm \
s}^{-1} \), for the two cases, respectively, in contrast with
$v_0= 10^7 {\rm cm \ s}^{-1}$. Therefore, the rate of spinning
down of a superfluid by virtue of random unpinning of its vortices
(from Eqs 2, 9, and 17) may as well be much less, up to 12 orders
of magnitudes, than that predicted in the vortex creep model.
Consequently, the spinning down of the superfluid in the crust of
a neutron star, through random unpinning of its pinned vortices,
would have no significant effects on the observable post-glitch
spin-down behavior of the star.

Further studies should indicate which of the two cases, {\em (i)}
or {\em (ii)}, is relevant (if at all) and thus determine the
radial creep rate appropriately. It has to be verified that a
torque may be imparted on a superfluid while a large fraction of
its vortices and/or a large part of each creeping vortex is
pinned. Microscopic description of the vortex motion should
indicate the extent to which any single vortex may deviate from
Eq.~6 and behave independently (if at all). Otherwise a
generalization of the idea of creeping of the fluxoids, in {\em
random directions}, to the vortices of a stationary rotating
superfluid would be moot. While Magnus force and/or thermal
activation (quantum tunnelling) could cause unpinning however the
subsequent radial motion of the released vortices might as well be
``truncated'', except in the presence of an external torque on the
superfluid.

This work was supported by a grant from the Research Committee of
Shiraz University.

\end{document}